\documentclass{elsart}

 \usepackage{epsfig}

\usepackage{amssymb}

\begin{document}

\begin{frontmatter}



\title{\large{\bf Entropic Studies of Cytoskeletal Motors Jamming}}


\author{C.M. Arizmendi$^{1,*}$, H.G.E. Hentschel$^2$, and F. Family$^2$}

\corauth[cor1]{Corresponding author. E-mail: arizmend@fi.mdp.edu.ar}
\address
{
$^1$Depto. de F\'{\i}sica,
Facultad de Ingenier\'{\i}a, \\
Universidad Nacional de  Mar del
Plata,\\
Av. J.B. Justo 4302, \\
7600 Mar del Plata,
Argentina\\
$^2$Department of Physics,
Emory University, \\
Atlanta, GA 30322,  USA
}

\begin{abstract}
Can the different causes for disruption of intracellular transport
 be  traced from the trajectories of the molecular motors on the cytoskeletal
 filaments? We will attempt to answer this important question in a
 Monte Carlo model of microtubule-motor protein interaction from
 the point of view of information theory.
 \vfill
\end{abstract}

\begin{keyword}
Molecular Motors \sep Intracellular transport diseases  \sep Data
Compression
\PACS  87.16.Nn  \sep 87.10.+e \sep 89.70.+c
\end{keyword}
\end{frontmatter}

Molecular motors (MM) are key means of intracellular transport and
therefore control the spatial organization of eukaryotic cells
\cite{motors}. One intriguing example is provided by the
intracellular traffic of MM along cytoskeletal filaments.
Microtubule-MM interaction play a crucial role in cellular
processes, such as chromosomal segregation during cell division,
flagellar motion and axonal transport. Disruption of transport
events has been determined to be the molecular basis for many
diseases like Alzheimer's disease and motor neuron disease (MND)
\cite{jam}.
 Large axonal
swellings, spheroids, in the spinal cords of patients with motor
neuron disease show massive accumulation of kinesin, major
molecular motors responsible for fast axonal transport
\cite{motodis}.

Disturbances of cytoskeletal filament transport has been also
studied recently from a theoretical point of view \cite{Lipo},
finding that hard-core motor-motor interactions produce
overcrowding of the motors on the filament near its $+$ end, even
at relatively small motor concentration, and this traffic jam then
extends towards the $-$ end with increasing concentration of
motors. This motor jamming produce also a decrease of the current
of motors through the filament. The average motor current obtained
as a function of the motor concentration is shown in Fig. 1. It
can be seen from that figure that a low average motor current may
be caused by low motor concentration or by high motor
concentration with jamming. Even in this simple model, the average
motor current is useless to distinguish the two possible causes of
current decrease. Can the different causes of current decrease
like overcrowding, low motor density, or other motility problems
like those of MND, be traced from the molecular motors
trajectories? The entropy of the motor walk seems one possibility
that we will study here.

 Molecular motors interacting with cytoskeletal filament  constitute
far from equilibrium, non-ergodic systems and the usual  entropy
definition depending  on probabilistic ensemble concepts are not
well suited to describing the information content of this
dynamics.  Kolmogorov's concept of algorithmic complexity
\cite{Li} is commonly used in the study of far from equilibrium
systems, such as proteins and fractal growth processes, because it
was developed to deal with single objects.
\begin{figure}[b]
{\psfig{file=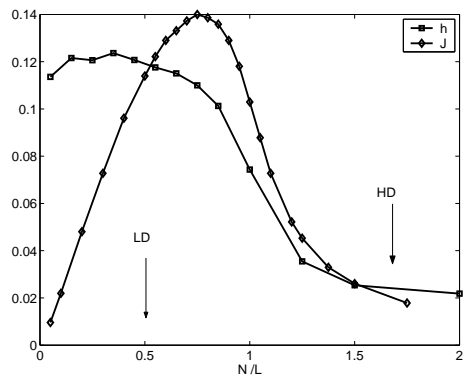,height=5cm}} \caption{Plots of the entropy per
character $h$, and average current $J$, as a function of the linear
density $N/L$ of motors.}
\end{figure}
The algorithmic complexity of an object is broadly defined as the
length in bits of the shortest description for that object
\cite{Li}. The algorithmic complexity of a string of characters is
the length in bits of the smallest program that produces as output
the string. The problem with this definition is that it is
impossible, even in principle, to find such a program.
Nevertheless, the zipping file compression programs are based on
algorithms that are expected to carry out this task, at least
approximately. The Lempel and Ziv algorithm ($LZ77$) \cite{LZ77},
used by {\sl gzip} and {\sl zip} is probably the best known file
compression algorithm. In this algorithm, the data sequence is
parsed into words such that the next word is the shortest word not
seen in the past. The coding of this new word is done by a pair of
numbers formed by the pointer to the last occurrence  of the
prefix and the last bit of the new word. Therefore the zipper will
encode more frequent sequences with fewer bytes and will use more
bytes for rare sequences. It has been shown that \cite{LZ77} when
the length of text to be zipped tends to infinity the ratio
between the length of the zipped file and the length of the
original file tends to $h$, the entropy per character of the
ergodic source of the text.  The $LZ77$ zipper was used recently
for language and author recognition\cite{tanos}, to quantify the
information transfer in thermal ratchets\cite{colonia}, and in
Parrondo's games\cite{Pau}.

We will address the motor dynamics from this point of view of
information theory. As the motors either move forward or remain at
a fixed site at a given time step, the trajectory of any given
motor while bound to the filament is isomorphic to a message
composed from an alphabet consisting of two letters. Thus our
procedure to analyze the bound motor dynamics is as follows: For a
given concentration of motors, every motor trajectory on the
microtubule is transformed on a binary message. All these messages
are joined into a single string, and the entropy per character of
this string is obtained with the zipping method
 as the ratio between the length of the zipped file and the length of the original file.
 The entropy  per character $h$ obtained in this way  is independent of the length of the
 string and corresponds to the properties of the
 ergodic source of the string of messages. Specifically $h$ will be the average number of
 bits of information that we gain on being told
 that a given motor
 moves forward or remains immobile at a given site. This procedure is carried out for different
 concentrations of motors.
We use a lattice model developed by Lipowsky, Klumpp and
Niewenhuizen \cite{Lipo} to represent the molecular motor traffic
along a microtubule. Here we focus on a closed cylindrical
compartment as in Fig. 1 of \cite{Lipo} with one microtubule in
the center. The motor particle moves
   on a cubic lattice with lattice constant $l$.
The cylindrical compartment has length $L = 200l$ and radius
$L_{\perp} = 25l$. Its $+$ end is at $x=L$.
 In order to
   perform MC simulations simulation parameters were
   chosen in
 such a way that the lattice random walks have
   the same bound state velocity, bound and unbound diffusion
   coefficient $D_b$ and $D_{ub}$,
 respectively, and walking
   time $\Delta t_b$ as the real motors. The motor particle moves
   on a cubic lattice with lattice constant $l$.
 The
   microtubule is taken to consist  of a one-dimensional
   line of binding sites. Away from the filament, the hopping
   rates between any
 two nearest neighbor sites are equal
   to $1/6 \tau$ where the time scale $\tau$ for the unbound
   motor is given by $\tau = l^2/6D_{ub}$.
 The motor can
   adsorb onto a filament binding site with sticking probability
   $\pi_{ad}$. Once the motor is bound, the time scale changes
    to $\tau_b$. $\alpha/\tau_b$ is the forward step rate and
    $\epsilon/\tau_b$ the unbinding rate. We ignore backward
    steps because we are
 simulating processive motors. Nevertheless, bound state diffusion still occurs because
    of the nonzero
     dwell
    probability defined by $\gamma = 1 - \alpha -
 2/3 \epsilon$.
     The dwell probability $\gamma$
     defines the mean dwell time
     $\tau_{dw} = \tau_b \gamma/(1-\gamma)$ and the mean step time
     $\tau_s \equiv \tau_{dw} + \tau_b$.
     In our MC simulations we
  focus on two headed kinesin
 \cite{Lipo} which is
  experimentally characterized by a filament repeat length
  $l= 8nm$, a bound state velocity $v_b = 680 nm/s$ and a bound
   state diffusion coefficient $D_b = 1360 nm/s^2$, which are recovered
 \cite{Lipo} for
      $\alpha = 0.4975$, $\gamma =  0.4987$, and
      $\epsilon = 0.0075$; $\tau_b = 5.9 ms$, $\tau_s = 11.8 ms$.
      The bulk
 diffusion coefficient for unbound motors is taken as
$D_{ub} = 4\mu m^2/s$ which implies $\tau = \tau_b/1341$, and
$\pi_{ad} = 1$. Within the compartment, the active transport of
the motors along the filament produces both, a bound current
$J_b$, and, a concentration gradient between the two ends of the
tube. While the unbound motors give rise to a diffusive current
$J_{ub}$ that balances the bound current along the filament
resulting in a nonequilibrium steady state distribution of both
bound and unbound motors in the axonal compartment.
 The  motion of every motor on the microtubule resulting from the MC simulation is
coded into a binary string where ${\bf 0}$ and ${\bf 1}$
correspond to no motion or a step to the $+$ end of the
microtubule respectively as described above. The unbound motor
motion is not taken into account.
 The binary strings obtained for different concentrations of motors were
 zipped to obtain approximations to  the entropy
per character $h$ for different motor concentrations. The overall
transport along the filament can be characterized by the average
current $J = \int dx J_b(x)/L$. The  average current $J$ and the
entropy per character $h$ as functions of the linear density of
motors $N/L$ are shown in Fig. 1. The  average current $J$
dependence on the linear density of motors $N/L$ is analogous to
the fundamental diagram of  traffic models \cite{Nagel} in which
the car flow as a function of vehicle density increases linearly
for low densities, reaches maximum throughput, and then decays for
higher densities. The analogy between both, Nagel's  traffic model
and our molecular motor model, shows clearly that the scenario is
the same for both models, and we may imply that the critical phase
transition at maximum throughput of Nagel's  traffic model
\cite{Nagel} is also present in our model for maximum MM average
current. Critical phase transition is associated to divergence of
correlation between motors and continuous entropy variation
\cite{Stanley} across the transition density point as can be seen
in Fig. 1.

 A low
average motor current may be caused by low motor concentration or
by high motor concentration with jamming. Nevertheless the entropy
per character $h$ is much higher for the low density (LD) free
motion than the corresponding to the high density (HD) jammed
regime.
 This behavior is due to the fact that for low $N$, motors move nearly free on the
 microtubule
 and dwell and step
forward probabilities, $\gamma$ and $\alpha$,  are almost the
same. On the other hand, for large $N$, $h$ decreases because
overcrowding limits the stepping forward of the motors. In this
way, the ${\bf 1}$ that indicates a step towards the $+$ end
becomes unusual and the strings are composed mainly by the
"motionless" ${\bf 0}$. According to Shannon information theory,
entropy measures {\it how much information is gained when an
outcome is observed}. Nothing new is learned when a usual ${\bf
0}$ is found, but ${\bf 1}$ at the string is a piece of
information that decreases the level of uncertainty.

 Another important quantity that may be
used is the notion of relative entropy or Kullback-Leibler
divergence \cite{Kullback} which is a measure of statistical
remoteness between two distributions. Recently Ziv and Merhav
\cite{Ziv} proposed an algorithm based on a procedure similar to
the one used in the $LZ77$. A nice review of Ziv and Merhav
algorithm with interesting applications was done in \cite{tanos2}.
We follow the recipe described in \cite{tanos2}: The relative
entropy between two distributions of MM trajectories corresponding
to different concentrations of motors ${\bf N_1}$ and  ${\bf N_2}$
that produce binary strings $N_1$ and $N_2$, respectively,  is
obtained in the following way: Long sequences $n_1$ and $n_2$ are
extracted from the binary strings $N_1$ and $N_2$, respectively,
and a short sequence $n_{2sh}$ from $N_2$. New sequences $n_1 +
n_{2sh}$ and $n_2 + n_{2sh}$ are created by appending $n_{2sh}$ to
$n_1$ and $n_2$, respectively. The relative entropy per character
$h_{N_1N_2}$ is estimated by
$$
h_{N_1N_2}= \frac{Z_{n_1 + n_{2sh}} - Z_{n_1} -(Z_{n_2 +
n_{2sh}}-Z_{n_2})}{|n_{2sh}|},
$$

where $Z$ is the length of the zipped sequence, and $|n_{2sh}|$ is
the number of characters of the sequence $n_{2sh}$. It may be
appreciated in Fig. 2 that relative entropy estimates distinguish
clearly between the strings that result from
 high and low motor density dynamics.

In summary, by mapping the molecular motor motion to strings of
characters, with data compression techniques we obtained the
change of entropy corresponding to the high-low density
transition. The overcrowding cause  can be distinguished from the
low  motor density cause of cytoskeletal motor current decrease
with the entropy per character $h$ and also by the relative
entropy per character $h_{N_1N_2}$. We will present elsewhere the
results for the motor neuron disease simulation.
\begin{figure}
{\psfig{file=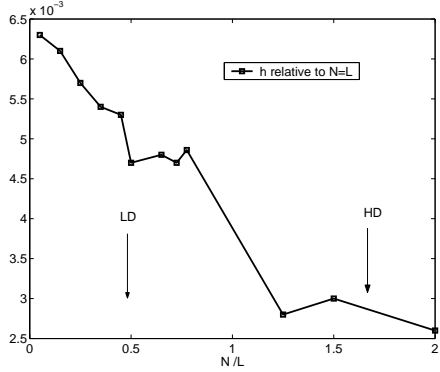,height=5cm}} \caption{Estimate of relative
entropy between different concentrations and $N=L$, approximate
limit between high and low motor density.}
\end{figure}

This work was partially supported by grants from UNMDP, ANPCyT
(PICTO11-090076) and by NSF Grant No. IBN-0083653. We thank
anonymous referees for helpful comments. C.M.A. wants to
acknowledge M.C. Azpiazu for talks on long delay times, random
walks and uncertainties in the web while this work was done.


\end{document}